\documentclass[fleqn,twoside,twocolumn,nofootinbib,showkeys]{revtex4} 
\usepackage[nocpr]{ujp} 

\begin{document}
\title[Calculation and Experimental Study]
{CALCULATION AND EXPERIMENTAL STUDY\\ OF THE RETRACTING FORCE FOR
MAGNETIC\\
SPRINGS OF TWO TYPES}
\author{V.Yu.~Tsivilitsin}
\affiliation{I.M.~Frantsevich Institute for Problems of Materials
Science, Nat. Acad. of Sci. of Ukraine}
\address{3, Krzhyzhanivskyi Str., Kyiv 03142, Ukraine}
\email{ibond@ipms.kiev.ua}
\author{Yu.V.~Milman}%
\affiliation{I.M.~Frantsevich Institute for Problems of Materials
Science, Nat. Acad. of Sci. of Ukraine}%
\address{3, Krzhyzhanivskyi Str., Kyiv 03142, Ukraine}%
\author{V.A.~Goncharuk}
\affiliation{I.M.~Frantsevich Institute for Problems of Materials
Science, Nat. Acad. of Sci. of Ukraine}
\address{3, Krzhyzhanivskyi Str., Kyiv 03142, Ukraine}
\author{I.B.~Bondar}%
\affiliation{I.M.~Frantsevich Institute for Problems of Materials
Science, Nat. Acad. of Sci. of Ukraine}%
\address{3, Krzhyzhanivskyi Str., Kyiv 03142, Ukraine}%
\email{ibond@ipms.kiev.ua} \udk{???}  \razd{}

\autorcol{V.Yu.\hspace*{0.7mm}~Tsivilitsin,
Yu.V.\hspace*{0.7mm}~Milman, V.A.\hspace*{0.7mm}~Goncharuk et al.}

\setcounter{page}{1020}%

\begin{abstract}
Designs for magnetic springs of two types have been proposed, and
the methods of calculation of their retracting forces have been
developed.\,\,Formulas are obtained for the retracting force in the
main section of spring force characteristics.\,\,Experimental data
are in good agreement with the results of theoretical calculations.
The force characteristics of the proposed magnetic spring
constructions can be varied for a specific application.\,\,The
derived formulas are verified experimentally.\,\,Ways to change the
force characteristics of magnetic springs according to specific
requirements are demonstrated.
\end{abstract}
 \maketitle

\section{Introduction}\vspace*{-0.5mm}

There are a lot of magnetic spring types.\,\,The majority of them
have a power-law dependence of the retracting force on the
displacement.\,\,However, the largest practical interest is
attracted by magnetic springs, whose force weakly depends on the
displacement \cite{1,2}.

In two last years, we studied in detail two types of magnetic
springs with an almost constant retracting force (the corresponding
variation varies within the limits of 15\%) over the working stroke
of the spring. We consider the springs of those types to be the most
promising for the majority of specific applications.

The first type of magnetic springs can be conditionally called
\textquotedblleft permanent magnet--magnetic
circuit\textquotedblright\ (Fig.~1).\,\,In this type of springs, the
interaction between a permanent magnet (as a rule, with a large
energy product) and a magnetic circuit fabricated from a
magnetically soft material is used.\,\,The second type of magnetic
springs can be conditionally called \textquotedblleft two permanent
magnets\textquotedblright\ (Fig.~2).\,\,It consists of a tubular
permanent magnet with the axial magnetization, with another
cylindrical or ring magnet inside, the magnetization of which is
antiparallel to that of the external magnet.\,\,This design allows
the force characteristic shape to be varied by applying of a soft
magnetic disk on the non-working end of the spring (see Section
\ref{sec4}).

In this work, the calculation technique is developed, and the
corresponding formulas are obtained for the determination of the
retracting force magnitude in the main section of force
characteristics, the models for magnetic springs of both types are
developed, and the methods to vary the shape of a force
characteristic at its initial and finite sections was
proposed.\,\,All obtained results are verified experimentally on an
automated installation for mechanical tests R-5.

\section{Physical Model and Derivation of the Retracting Force Formula for
a Magnetic Spring of the First Type}

In Fig.~1, the schematic diagram of a magnetic spring of the first
type (\textquotedblleft permanent magnet--magnetic
circuit\textquotedblright) is shown.\,\,The permanent magnet can be
a cylinder with diametral magnetization or a rectangular prism with
magnetization directed perpendicularly to a displacement.

The cross-section of the magnet oriented perpendicularly to a spring
displacement direction can also have the shape of ellipse, rhombus,
or trapeze: in all cases, the spring will work, but with a weaker
retracting force, because the demagnetizing factor is smaller in
those cases.\,\,The shape of the magnetic circuit must satisfy the
following requirement: it must contain a cavity with the transverse
cross-section that corresponds to that of the permanent magnet (it
is desirable that the magnet should enter it with a sli\-ding~fit).

For the spring to be the most efficient, the magnetic circuit should
transmit the total magnetic flux created by the permanent
magnet.\,\,The smaller the gap between the magnet poles and the
magnetic circuit, the lower is the magnetic flux scattering, and the
stronger is the retracting force of the magnetic spring. The
distance between the lateral sides (not poles) of the permanent
magnet and the magnetic circuit can be arbitrary; it practically
does not affect the force characteristics of magnetic springs of
this type.

Earlier, we showed \cite{3} that the retracting force can be calculated by the
formula:%
\begin{equation}
F=N^{2}(B_{r}^{2}/2\mu_{_{0}})S,   \label{eq1}
\end{equation}
where $N$ is the demagnetizing factor dependent only on the
geometric shape of the magnet, $B_{r}$ the residual magnetic
induction of the material that the permanent magnet is fabricated
from, and $S$ the area of a transverse cross-section of the
permanent magnet.\,\,Formula (\ref{eq1}) was obtained, by using the
virtual displacement method.\,\,At retracting the magnet by the
distance $dx,$ the work
\begin{equation}
dA=Fdx   \label{eq2}
\end{equation}
is executed, where $F$ is the retracting force.\,\,On the other
hand, before the permanent magnet was introduced into the magnetic
circuit, the demagnetizing field \cite{4}
\begin{equation}
B=-NB_{r}   \label{eq3}
\end{equation}
existed inside the magnet owing to the existence of magnet
poles.\,\,In this expression, $N$ is the demagnetizing factor, and
$B_{r}$ the residual induction of a magnetically hard material.
After the magnet is introduced into the magnetic circuit, this field
almost vanishes, so that the demagnetizing factor can also be
conditionally adopted to equal zero.\,\,Therefore, in our
approximation, the whole energy of the demagnetizing field can be
considered to be spent on the work associated with a displacement of
the permanent magnet.\,\,The magnetic field energy density equals
\begin{equation}
w=B^{2}/2\mu_{_{0}}=N^{2}B_{r}^{2}/2\mu_{_{0}},   \label{eq4}
\end{equation}
where $\mu_{0}=4\pi\times10^{-7}~\mathrm{H/m}$ is the universal magnetic
constant.

\begin{figure}%
\vskip1mm
\includegraphics[width=6cm]{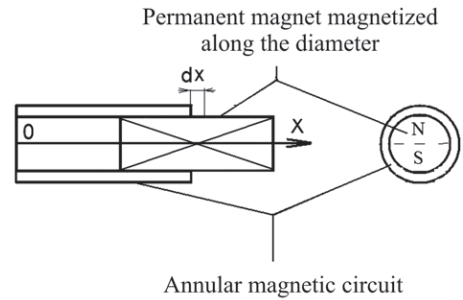}
\vskip-4mm\caption{Schematic diagram of the magnetic spring with a
cylindrical magnet with diametral magnetization   }
\end{figure}

If the magnet shifts by the distance $dx$, the magnetic field energy $W$
changes by
\begin{equation}
dW=wdV,   \label{eq5}
\end{equation}
where $w$ is the density of the magnetic field energy, and $V$ the
volume, in which the field has changed.\,\,In our case,
\begin{equation}
dV=Sdx,   \label{eq6}
\end{equation}
where $S$ is the area of a magnet cross-section perpendicular to the
cylinder axis.\,\,Substituting Eqs.~(\ref{eq4}) and (\ref{eq6}) into
Eq.~(\ref{eq5}), we obtain
\begin{equation}
dW=N^{2}(B_{r}^{2}/2\mu_{_{0}})Sdx.\,\,\label{7}
\end{equation}
Equating Eqs.~(\ref{eq2}) \textrm{and} (\ref{7}), i.e.\,\,putting
$dA=dW$, and reducing the result by $dx$, we obtain the simple
expression (\ref{eq1}) for the force $F$ of the permanent magnet
retraction into a magnetic circuit.

Formula (\ref{eq1}) for the retracting force is in good agreement
with experimental data.\,\,A typical experimental force
characteristic of such a spring is depicted in Fig.~3.\,\,The
difference between the force characteristics obtained at the
retraction and the pulling is connected with the friction force
between the permanent magnet and the magnetic circuit: the friction
force is added to the retracting force at the pulling (upper curve),
and subtracted at the retraction (lower curve).\,\,This design of a
magnetic spring was protected by a patent for utility model
\cite{5}.

We also experimentally proved the strong influence of the
demagnetizing factor $N $ on the retracting force \cite{3} for
magnetic springs with permanent magnets in the form of rectangular
parallelepipeds characterized by the identical area of transverse
cross-sections, but different distances between the poles.\,\,In
accordance with formula (\ref{eq1}), the forces turned out different
by a factor of 1.7.

\begin{figure}%
\vskip1mm
\includegraphics[width=7cm]{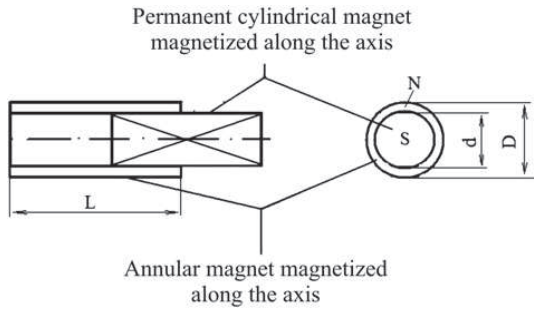}
\vskip-3mm\caption{Schematic diagram of a magnetic spring
\textquotedblleft two permanent magnets\textquotedblright: $D$ and
$d$ are the external diameters of the ring and cylindrical,
respectively, magnets,; $L$ the length of the external magnet}
\end{figure}
\begin{figure}%
\vskip3mm
\includegraphics[width=7cm]{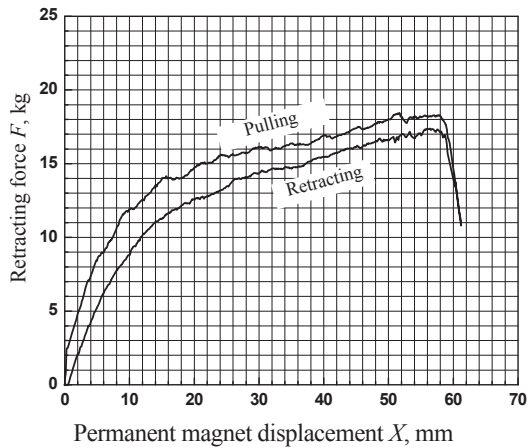}
\vskip-3mm\caption{Dependence of the retracting force on the
displacement of the magnet in the tubular magnetic circuit for the
cylindrical magnet 30~mm in diameter fabricated from Nd--Fe--B
alloy, with diametral magnetization and the residual induction
$B_{r}=$ $=1.25~\mathrm{T}$}
\end{figure}

The advantages of this magnetic spring type include an easily
controllable stroke length.\,\,The stroke length of a spring depends
on the length of an applied magnet.\,\,For a larger stroke, a longer
magnet should be used.\,\,In this case, the retracting force does
not change.\,\,As a shortcoming of this design, we can consider a
strong force of attraction of the permanent magnet poles to the
magnetic circuit.\,\,It can amount to 30--40\% of the retracting
force if the magnet is badly centered in the magnetic
circuit.\,\,Both magnet poles should be equidistant from the
magnetic circuit in order to reduce the friction force, which can
result in an appreciable hysteresis in the spring force
characteristic (Fig.~3).

In our experiments, we used ground magnets, which enter with a
sliding fit into the magnetic circuit fabricated from E12
electrotechnical steel.\,\,In order to reduce the friction force,
the thin layer of a non-magnetic material with low friction
coefficient can be used.\,\,This layer can also serve as an
anticorrosion protection for the permanent magnet from humid air and
help to mount the magnet poles at the same distance from the
magnetic circuit.\vspace*{-2mm}

\section{Physical Model and Derivation of the Retracting Force Formula for
a Magnetic Spring of the Second Type}

In our opinion, another design of a magnetic spring, which can be
called \textquotedblleft two permanent magnets\textquotedblright ,
has wider capabilities.\,\,In the construction of this type, an
external ring magnet and an internal cylindrical magnet are
used.\,\,As the latter, a magnet with more complicated shape, but
with cylindrical symmetry, e.g., a truncated cone, can be
used.\,\,The magnetizations of both magnets are axial and
antiparallel.\,\,The schematic diagram of such a spring is depicted
in Fig.~2.

The internal magnet with a more complicated shape can be useful if
the modification of the force characteristic of a magnetic spring is
desirable. This construction can include a thin non-magnetic ring
between the magnets to provide required changes in the force
characteristic of the spring.\,\,The ring diminishes a little the
retracting force in the main section, because the transverse
cross-section area of the magnet becomes smaller.\,\,The
corresponding decrease can be easily estimated, by using the formula
derived below.

The retracting force can be varied by decreasing the area of the
internal magnet pole.\,\,One can either diminish the internal magnet
diameter or make an axial aperture in the latter, which turns out
very useful if the magnet is attached to the control rod.\,\,The
matter is that permanent magnets of the Nd--Fe--B system are
characterized by the heated steel hardness, being at the same time
very fragile, which makes it almost impossible to create any carving
on them.

Let us consider a detailed derivation of the formula describing the
retracting force for the design of this type.\,\,Earlier, it was
shown \cite{6}
that the retracting force is well described in this case by the formula%
\begin{equation}
F=B_{r}HS/\mu _{_{0}},  \label{eq7}
\end{equation}%
where $S$ is the area of the transverse cross-section of the internal magnet, $%
B_{r}$ the residual magnetic induction of a material, which the permanent
magnet is fabricated from, $\mu _{0}$ the universal magnetic constant, and $H
$ the field in the external ring magnet, which is determined by the formula
\cite{4}%
\begin{equation}
H=B_{r}[(1+D^{2}/L^{2})^{-0.5}-(1+d^{2}/L^{2})^{-0.5}].\,\,\label{eq8}
\end{equation}%
Here, $D$ and $d$ are the external and internal, respectively, diameters of
the tubular magnet, and $L$ its length.

If the residual inductions of both permanent magnets are identical,
we obtain from Eqs.~(\ref{eq7}) and (\ref{eq8}) that
\[
F=B_r ^2[(1+D^2/L^2)^{-0.5}\,-\]\vspace*{-8mm}
\begin{equation}
\label{11} -\,(1+d^2/L^2)^{-0.5}]S/\mu _0 .
\end{equation}
In the case concerned,%
\begin{equation}
S=\pi d^{2}/4.  \label{12}
\end{equation}%
Substituting Eq.~(\ref{12}) into Eq.~(\ref{11}), we have%
\[  F=B_r ^2[(1+D^2/L^2)^{-0{.}5}\,-\]\vspace*{-8mm}
\begin{equation}
\label{eq9} -\,(1+d^2/L^2)^{-0{.}5}]\pi d^2/4\mu _0.
\end{equation}
Hence, we obtained the formula for the dependence of the retracting
force magnitude on the geometric dimensions of applied magnets for
a magnetic spring of the second type.

A comparison of magnetic springs of both types with identical
diameters demonstrates that the second spring provides a retracting
force, which is 40--100\% stronger.\,\,Another advantage of this
magnetic spring consists in an insignificant attraction between the
internal and external magnets (no substantial hysteresis in the
force characteristic resulting from the friction force was detected
experimentally).\,\,A typical experimental force
characteristic of a magnetic spring of the second type with $D=40$~mm and $%
L=35$~mm is shown in Fig.~4.\,\,This design of a magnetic spring
also obtained the patent of Ukraine for utility model \cite{7}.

Perhaps, the only shortcoming of this construction is the increase
of the spring stroke length, which is not so simple as in the first
case.\,\,The matter is that if the stroke length grows, but the
external diameter of the spring remains constant, the retracting
force considerably decreases in accordance with formula (\ref{eq9}).
Now, let us consider its advantages.

\begin{figure}%
\vskip1mm
\includegraphics[width=7cm]{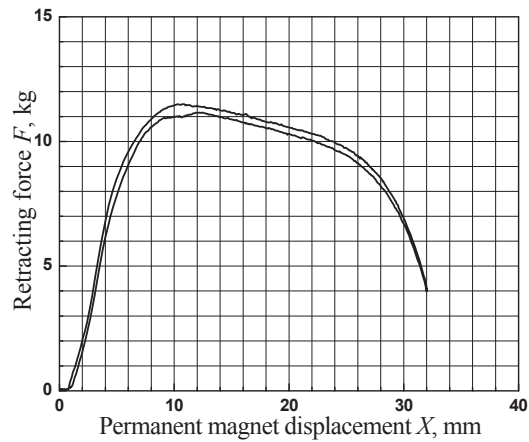}
\vskip-3mm\caption{Dependence of the retracting force $F$ on the
core displacement $X$ for a magnetic cylindrical spring with the
following parameters: the external diameter of the tubular magnet
equals 40~mm, the internal diameter of the tubular magnet 20~mm, the
magnet length 30 mm, the internal magnet length 35~mm, and the
internal magnet diameter 19.7~mm}
\end{figure}

1.~As was said above, a magnetic spring of the second type is
40--100\% more powerful provided the same diameter of the external
magnet.

2.~It can be optimized to a specific application depending on which
parameter is a key one in this application.\,\,For instance, the
diameters of magnets can be calculated proceeding from the stroke
length of the spring and the required operating force.

3.~It is easily to provide a latch effort, which can be 2 to 4
times as large as the spring operating force, and the operating
parameters of the spring in the main section will not be changed at
that.

 4.~The section, where the spring force increases (in Fig.~4, these
displacements are less than 10~mm) can be almost completely excluded.

5.~The friction force between the moving magnets is insignificant,
and the exact centering of magnets, which is required in the case of
a magnetic spring of the first type, is not necessary.

\section{Magnetic Spring of the Second Type\\ with a Soft Magnetic Disk on Non-Working\\ End of the Spring}

\begin{figure}%
\vskip1mm
\includegraphics[width=7cm]{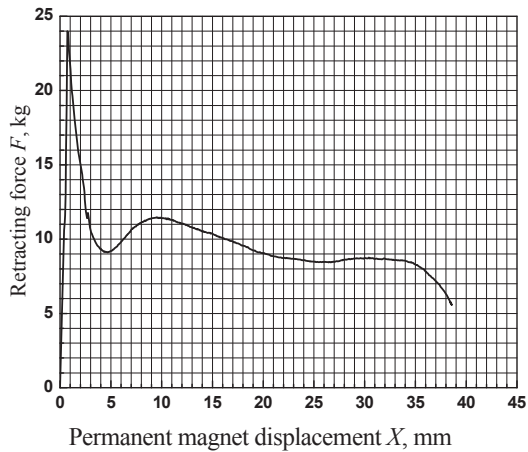}
\vskip-3mm\caption{Force characteristic of a spring of the second
type with the end magnetic circuit: the external diameter of a
tubular magnet equals 40~mm, the internal diameter of a tubular
magnet 28~mm, the internal magnet length 35~mm, and the internal
magnet diameter 25~mm}
\end{figure}
\begin{figure}%
\vskip3mm
\includegraphics[width=7cm]{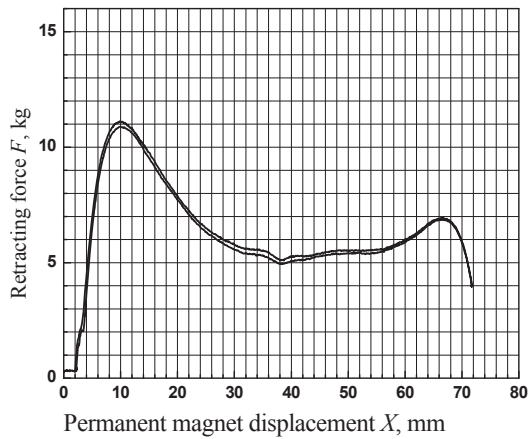}
\vskip-3mm\caption{Force characteristic of a magnetic spring with
the double length: the tubular magnet length equals 70~mm, the
external diameter 40~mm, and the internal diameter 25 mm}
\end{figure}

The construction of the second type can be easily used, e.g., for
the development of a door closer with prolonged service
life.\,\,From this point of view, it satisfies all requirements put
to those units: a retracting force of about 100--150~N in the main
section and a required latch effort for the complete door closing.
The latch effort is provided by attaching a cylindrical soft
magnetic disk to the inactive end of the spring.\,\,The typical
force characteristic of such a magnetic spring with a soft magnetic
disk on the non-working end of the spring is depicted in
Fig.~5.\,\,To make the latch effort stronger, the right end of the
internal magnet can also be connected with a soft magnetic disk.

Any magnitude of the latch effort can be obtained within the
interval from zero to the maximum by introducing a non-magnetic
interlayer with a required thickness between the soft magnetic disk
and the internal magnet.\,\,This means that the curve in Fig.~5 can
begin from any point within the interval from zero to 25~kg on the
ordinate axis.\,\,The latch effort can be controlled by varying the
thickness of the end magnetic circuit.

This design was successfully used by us in the return valve applied
at the oil well washing.\,\,The valve is actuated when the pressure
increases to 40~atm and is closed when the pressure drops below
10~atm.\,\,It is very difficult, if possible at all, to provide such
a force characteristic with the help of ordinary mechanical
springs.\vspace*{-2mm}

\section{Influence of Magnet\\ Ends on the Force Characteristic}

\label{sec4}

In the magnet spring construction of the second type, a large role
in the formation of a force characteristic is played by the magnet
ends, with their influence increasing with the ratio between the
spring length and the spring diameter.\,\,A typical force
characteristic of the spring usually contains two maxima (see
Fig.~6).\,\,The first maximum is always larger than the second,
because it is formed by both magnet poles.\,\,The \textquotedblleft
plateau\textquotedblright\ in the exhibited force characteristic
(the section from 30 to 60~mm) can be calculated with a sufficient
accuracy by formula (\ref{eq9}), and the maxima are a consequence of
the magnet end action.\,\,We experimentally proved this statement
using a spring with the same diameter, but a double length.\,\,As a
result, the corresponding maxima became sharper, and the
\textquotedblleft plateau\textquotedblright\ diminishes to an effort
of 5.5~kg (Fig.~6).

If a smoother force characteristic of the spring is required, those
maxima can be \textquotedblleft cut off\textquotedblright\ by
reducing the area of end cross-sections.\,\,Substituting the central
cylindrical magnet by a combined one consisting of two truncated
cones with a common base, it is possible to obtain a long enough
spring with almost constant retracting force.\vspace*{-2mm}

\section{Conclusions}

1.\,\,Designs of magnetic springs of two types--\textquotedblleft
magnet--magnetic circuit\textquotedblright\ and \textquotedblleft
two permanent magnets\textquotedblright--are proposed.\,\,Each of
them is protected by the patent of Ukraine \cite{5,7}.

2.\,\,Formulas for the dependence of the retracting force in the
main section of a displacement are derived.\,\,They are confirmed by
experiments.

3.\,\,On the basis of the models proposed, the parameters of
magnetic springs can be calculated, e.g., for the required spring
stroke length and effort.

4.\,\,On the basis of the obtained formulas, the suitability of the
proposed constructions for their specific applications can be
estimated.

5.\,\,The optimization of a magnetic spring like \textquotedblleft
Which geometrical dimensions of the magnetic spring are required to
provide the maximum ret\-rac\-ting force\textquotedblright\ or
\textquotedblleft Which type of magnetic springs with the same
retracting force requires the minimum of a mag\-ne\-tic substance at
its fabrication?\textquotedblright\ can also be~done.

6.\,\,The design \textquotedblleft two permanent
magnets\textquotedblright\ allows one to obtain very diverse shapes
of the force characteristic owing to the application of magnetic
circuits, which satisfy almost any specific requirements.

\vspace*{-5mm}
\rezume{%
В.Ю. Цивіліцин,\\ Ю.В.~Мільман, В.А.~Гончарук,
І.Б.~Бондар}{РОЗРАХУНОК ТА ЕКСПЕРИМЕНТАЛЬНЕ\\ ДОСЛІДЖЕННЯ СИЛИ
ВТЯГУВАННЯ МАГНІТНИХ\\ ПРУЖИН ДВОХ ТИПІВ} {Запропоновано конструкції
магнітних пружин двох типів, розроблено методи розрахунку їх сили
втягування. Отримано формули для підрахунку сили втягування на
основній ділянці силової характеристики пружин. Експериментальні
дані знаходяться у добрій відповідності до теоретичних розрахунків.
Силові характеристики запропонованих конструкцій магнітних пружин
можна змінювати залежно від конкретного застосування. Проведено
експериментальну перевірку отриманих формул. Показано шляхи зміни
силових характеристик магнітних пружин відповідно до вимог
конкретних застосувань.}

\end{document}